\documentclass[preprint2]{aastex631}
\usepackage{amsmath}

\newcommand{\oiii}{[O\thinspace {\sc{iii}}]}

\newcommand{\nii}{[N\thinspace {\sc{ii}}]}

\begin{document}

\title{The Star-Forming Main Sequence in JADES and CEERS at $z>1.4$: Investigating the Burstiness of Star Formation}

\author[0000-0003-1249-6392]{Leonardo Clarke}
\affiliation{University of California, Los Angeles, 475 Portola Plaza, Los Angeles, CA, 90095, USA}

\author[0000-0003-3509-4855]{Alice E. Shapley}
\affiliation{University of California, Los Angeles, 475 Portola Plaza, Los Angeles, CA, 90095, USA}

\author[0000-0003-4792-9119]{Ryan L. Sanders}
\altaffiliation{NHFP Hubble Fellow}
\affiliation{University of Kentucky, 506 Library Drive, Lexington, KY, 40506, USA}

\author[0000-0001-8426-1141]{Michael W. Topping}
\affiliation{Steward Observatory, University of Arizona, 933 N Cherry Avenue, Tucson, AZ 85721, USA}

\author[0000-0003-2680-005X]{Gabriel B. Brammer}
\affiliation{Cosmic Dawn Center (DAWN), Denmark}
\affiliation{Niels Bohr Institute, University of Copenhagen, Jagtvej 128, DK-2200 Copenhagen N, Denmark}

\author{Trinity Bento}
\affiliation{University of California, Los Angeles, 475 Portola Plaza, Los Angeles, CA, 90095, USA}

\author[0000-0001-9687-4973]{Naveen A. Reddy}
\affiliation{Department of Physics \& Astronomy, University of California, Riverside, 900 University Avenue, Riverside, CA 92521, USA}

\author{Emily Kehoe}
\affiliation{University of California, Los Angeles, 475 Portola Plaza, Los Angeles, CA, 90095, USA}



\begin{abstract}

We have used public JWST/NIRSpec and JWST/NIRCam observations from the CEERS and JADES surveys in order to analyze the star-forming main sequence (SFMS) over the redshift range $1.4 \leq z < 7$. We calculate the star-formation rates (SFRs) of the galaxy sample using three approaches: Balmer line luminosity, spectral energy distribution (SED) fitting, and UV luminosity. We find a larger degree of scatter about the SFMS using the Balmer-based SFRs compared to the UV-based SFRs. Because these SFR indicators are sensitive to star formation on different time scales, the difference in scatter may be evidence of bursty star-formation histories in the early universe. We additionally compare the H$\alpha$-to-UV luminosity ratio (L(H$\alpha$)/$\nu$L$_{\nu,1600}$) for individual galaxies in the sample and find that 29\%$-$52\% of the ratios across the sample are poorly described by predictions from a smooth star-formation history. Measuring the burstiness of star formation in the early universe has multiple significant implications, such as deriving accurate physical parameters from SED fitting, explaining the evolution of the UV luminosity function, and providing constraints for sub-grid models of feedback in simulations of galaxy formation and evolution.

\end{abstract}



\section{Introduction} \label{sec:intro}

The galaxy star-forming main sequence (SFMS) is a relation that exists between the stellar masses and the star-fomation rates (SFRs) of galaxies. This relation was first noted in a large sample of galaxies from the Sloan Digital Sky Survey \citep[SDSS,][]{2000AJ....120.1579Y,2004MNRAS.351.1151B}, and its evolution as a function of redshift has also been studied extensively \citep[e.g.,][]{2007ApJ...660L..43N,2012ApJ...754L..29W,2014ApJ...795..104W,2014ApJS..214...15S,2015ApJ...815...98S, 2017ApJ...847...76S}. Studies of the galaxy SFMS are typically aimed at constraining galaxy star-formation histories (SFHs) over time, and the tightness of the SFMS implies that galaxies of similar masses generally have similar SFHs characterized by a smooth buildup of stellar mass over time.

The intrinsic scatter in the SFMS is thought to encapsulate information about the stochasticity or ``burstiness" of galaxy SFHs. State-of-the-art hydrodynamical simulations of galaxy formation such as FIRE-2 \citep{2018MNRAS.480..800H}, Illustris TNG \citep{2018MNRAS.475..676S}, and EAGLE \citep{2015MNRAS.446..521S} make predictions of galaxy SFHs that depend on sub-grid models of feedback and star formation. Comparing observational constraints on galaxy SFHs with simulation predictions that are based on sub-grid models provides an opportunity to test and improve upon our understanding of how galaxies build up their stellar mass over time.

In addition to constraining simulations, bursty SFHs have been proposed as a possible explanation for the excess of galaxies at the bright end of the UV luminosity function at $z \gtrsim 10$ \citep[e.g.,][]{2023MNRAS.521..497M,2023MNRAS.519..843M,2023MNRAS.526.2665S,2023ApJ...955L..35S,2023MNRAS.525.3254S}. In short, the burstiness of galaxies may upscatter galaxies of fixed stellar mass to high UV luminosities, resulting in an excess of UV-bright galaxies at early times. Characterizing the burstiness of SFHs at early times is also important for SED-based determinations of stellar mass and SFR \citep[e.g.][]{2020ApJ...904...33L,2024MNRAS.530L...7H}. It has been shown, for example, that assuming a constant and/or smooth SFH during SED fitting can underestimate the stellar masses of star-forming galaxies on the order of $\sim$0.1-0.8 dex \citep{2015MNRAS.451..839D,2023arXiv230605295E,2023MNRAS.519.5859W,2024ApJ...963...74W}. Additionally, biases can be introduced into SED-fitting procedures due to recent star-formation episodes outshining older stellar populations \citep[e.g.,][]{2005ApJ...626..698S,2024ApJ...961...73N}.

There are several observational signatures that give insight into the SFHs of galaxies such as the intrinsic scatter in the SFMS, the H$\alpha$-to-UV ratio, and the deviation in log(SFR) from the SFMS \citep[e.g.,][]{2012ApJ...744...44W,2022MNRAS.511.4464A,2024MNRAS.52711372A}. A larger intrinsic scatter, for example, may indicate that galaxies undergo significant excursions from the SFMS over the course of their growth in the form of frequent bursts. Similarly, H$\alpha$ and UV emission are sensitive to SFRs on timescales of $\sim$5 Myr and $\sim$100 Myr, respectively, and their ratio can constrain the rate of fluctuations in the SFR. These observational signatures serve as metrics that can be directly compared with theoretical predictions of SFHs, thereby placing constraints on the proper implementation of sub-grid models of feedback and star formation. 

There is a large body of work that has compared observational signatures of bursty SFHs with theoretical predictions for galaxies at various redshifts \citep[e.g.,][]{2012ApJ...744...44W,2015MNRAS.451..839D,2019ApJ...881...71E, 2024MNRAS.52711372A}. It has been found, for example, that there may be a larger degree of intrinsic scatter in the SFMS for galaxies at low stellar masses ($\log({\rm M}_*/{\rm M}_{\odot}) \lesssim 8$), though a number of observational studies find no mass dependence in the intrinsic scatter \citep[e.g.,][]{2012ApJ...754L..29W,2015ApJ...799..183S,2015A&A...575A..74S,2016ApJ...820L...1K}. However, many of these previous observational studies rely solely on photometric data which probes star formation on $>$100 Myr time scales, or measure H$\alpha$ using photometric excesses which can be less certain when compared with direct spectroscopic measurement. With the increased spectroscopic sensitivity and wavelength coverage of the Near-Infrared Spectrograph \citep[NIRSpec;][]{2022A&A...661A..80J} on the {\it James Webb Space Telescope} ({\it JWST}), studies of the SFMS can now be done routinely for representative samples of high-redshift galaxies \citep[e.g.,][]{2023arXiv230602470L,2024MNRAS.527.6139S,2024ApJ...961...73N,2023arXiv230915720C,2024A&A...684A..75C}.

In this study, we use public \textit{JWST} data from the Near-Infrared Camera \citep[NIRCam;][]{2023PASP..135b8001R} and \textit{JWST}/NIRSpec observations from the JWST Advanced Deep Extragalactic Survey (JADES) and the Cosmic Evolution Early Release Science (CEERS) survey in conjunction with 3DHST photometric observations to measure the SFMS and its intrinsic scatter for galaxies at $1.4 < z < 7$. We measure SFRs using three different indicators: H$\alpha$ luminosity, UV luminosity, and SED fitting. The use of different indicators allows us to measure the intrinsic scatter in the SFR on different time scales, and therefore gain insights into the burstiness of galaxy star-formation histories in the early universe.

Throughout this study, we adopt $H_0 = 70\ {\rm km\ s^{-1}\ Mpc}$, $\Omega_m = 0.3$, and $\Omega_{\Lambda} = 0.7$ as cosmological parameters, and we assume a \citet{2003PASP..115..763C} initial mass function. Additionally, we use the solar abundances reported by \citet{2009ARA&A..47..481A} with $12+\log({\rm O/H})_{\odot} = 8.69$ and the solar metallicity as $Z_{\odot} = 0.014$.

\section{Observations and Measurements} \label{sec:style}

\subsection{Observations}
\subsubsection{CEERS}
%The CEERS data were reduced according to ... et al. (2023).
We use public NIRSpec data from the CEERS program \citep[Program ID: 1345][]{https://doi.org/10.17909/z7p0-8481,2023ApJ...946L..13F, 2023ApJ...951L..22A}. Our analysis is based on 6 NIRSpec Micro-Shutter Assembly (MSA) pointings in the AEGIS field, which utilized the grating/filter combinations of G140M/F100LP,
G235M/F170LP, and G395M/F290LP, providing a spectral resolution of $R\sim 1000$
over the wavelength range approximately $1-5\mu$m. Each pointing  was observed
for a total of 3107 sec in each grating/filter combination.  A 3-point nod pattern was adopted for
each observation along each 3-shutter MSA slit. In total, the 6 MSA pointings contained 321 slits and 318 distinct targets.
 
\subsubsection{JADES}

In addition to CEERS, we made use of publicly released data from the JADES program \citep[Program ID: 1210,][]{2023arXiv230602465E,https://doi.org/10.17909/8tdj-8n28} in the GOODS-S extragalactic legacy field. The full data release consisted of NIRCam imaging in multiple photometric bands as well as NIRSpec spectra taken with the PRISM/CLEAR, G140M/F070LP, G235M/F170LP, G395M/F290LP, and G395H/F290LP grating/filter configurations. Throughout this analysis, we utilized NIRSpec observations in the G140M/F070LP, G235M/F170LP, and G395M/F290LP grating/filter combinations.

Observations were conducted over the course of three different visits. In order to reduce noise introduced by detector defects, the MSA configuration was shifted to place targets on different slits for each visit. A total of 198 galaxies were observed with the aforementioned grating/filter configuration, 117 of which we were able to measure robust redshifts. Of this subset of 117 galaxies considered in this study, 37 were observed in all three visits, 25 were observed in two visits, and 55 were only observed during a single visit. For each visit, the exposure time was 2.3 hr in each grating, resulting in a total exposure time of 6.9 hr per grating for objects observed in all three visits. A more detailed description of the JADES NIRSpec observations can be found in \citet{2023arXiv230602467B}.

\subsection{Data Reduction}
\subsubsection{CEERS}
The CEERS data reduction was performed using the STScI \texttt{jwst} pipeline.\footnote{https://jwst-pipeline.readthedocs.io/en/latest/index.html} The details of the data reduction process can be found in \citet{2023ApJ...955...54S}. In short, the data were reduced using the pipeline, and exposures were co-added to produce a set of 310 individual 2D spectra. 1D spectra were then extracted from the 2D spectra according to the procedure described in Section \ref{sec:JADES_redux}, resulting in a sample of 252 1D spectra. These spectra were then corrected for wavelength-dependent slit losses \citep[see][]{2023ApJ...948...83R} and scaled so that the flux densities matched the broadband NIRCam photometry (or the 3DHST photometry where NIRCam imaging wasn't available).

\subsubsection{JADES} \label{sec:JADES_redux}
We downloaded reduced, two-dimensional (2D) NIRSpec spectra from the DAWN JWST Archive (DJA)\footnote{https://dawn-cph.github.io/dja/spectroscopy/nirspec/} whose reduction closely resembles the reduction procedure applied to the CEERS NIRSpec data. The reduction was performed using a custom data reduction pipeline, \texttt{MsaExp}\footnote{https://github.com/gbrammer/msaexp}, which takes the Stage 2 data products from the MAST database\footnote{https://archive.stsci.edu/hlsp/jades} as inputs and performs the wavelength calibrations, flat-fielding, photometric calibrations, and a point-source slit-loss correction on each of the individual exposures. Each of the 2D exposures was then co-added and combined into a final 2D spectrum for each object. A more detailed explanation of the reduction of the 2D spectra can be found in \citet{2023arXiv230600647H} and \citet{2024arXiv240402211H}.

The one-dimensional (1D) spectra were extracted from the 2D spectra by hand using an optimal extraction aperture \citep{1986PASP...98..609H} consistent with the procedure described in \citet{2023ApJ...955...54S}. During the extraction procedure, strong emission lines were identified by eye in the 2D spectra and were fit with a Gaussian profile to make initial estimates of their redshifts. 

Because the exposures were taken over multiple visits, there were uncertain variations in the position of the target on the slit during each visit. Since the exposures in each of these visits were combined into a single 2D frame, deriving a single solution for wavelength-dependent slit losses, accounting for the extended nature of galaxy targets, became non-trivial. However, we calculate that the slit-loss correction factors vary less than 5\% between $1-5\ \mu m$. Thus, we do not apply slit-loss corrections to the JADES spectra beyond the default point-source correction. We instead account for extended-source slit losses during the flux calibration stage.

In order to obtain a robust absolute flux calibration, we first ensured that the relative flux calibration between gratings was accurate before scaling the spectra in all gratings to match the available photometry. To ensure an accurate relative flux calibration, we applied a multiplicative scaling factor to the spectra in both the G140M and the G395M gratings, leaving the spectra in the G235M grating fixed. The relative scaling was applied under the assumption that all targets with the same exposure time should reach the same relative sensitivity between gratings. In other words, the median value of the error spectrum in the G140M grating ($med(\sigma_{G140M})$) relative to that for the G235M grating ($med(\sigma_{G395M})$) should be the same for all targets with the same exposure time. The same should be true for $med(\sigma_{G395M})$ relative to $med(\sigma_{G235M})$. Thus, we scaled each spectrum in the G140M grating such that the value of $med(\sigma_{G140M})/med(\sigma_{G235M})$ for each target was equal to the median value of $med(\sigma_{G140M})/med(\sigma_{G235M})$ across the whole sample. The same procedure was used to scale the spectra in the G395M grating relative to the G235M grating.

Subsequently, we produced synthetic photometric observations of all targets by passing the spectra through six JWST filter curves (F090W, F115W, F200W, F277W, F356W, and F444W). We then compared the synthetic photometry to the observed photometry and calculated a scaling factor in each filter. For the targets in which both the synthetic and the observed photometry were measured with a signal-to-noise ratio $>5$, we applied the median value of the scaling factors in all signicicantly detected filters to the spectra in all three gratings. This was done to each individual target where there were available photometric observations. Of the 117 spectra for which an initial redshift was determined during extraction, 56 targets had no absolute flux scaling applied due to lack of photometric observations or low signal-to-noise photometry. However, for the 61 objects with significantly detected photometry, the median absolute scaling factor is 1.405, and the standard deviation is 1.481 (0.17 dex), so we do not apply a multiplicative scaling factor to those 56 spectra.

\subsection{SED and Emission-Line Fitting}

The spectral energy distribution (SED) of each galaxy was fit using the full suite of available photometry from JWST/NIRCam, HST/WFC3, Spitzer/IRAC, and ground-based surveys to cover the rest-frame UV to the rest-frame infrared. 

For both the CEERS and JADES samples, we used publicly available multi-wavelength photometric catalogs constructed by G. Brammer\footnote{https://s3.amazonaws.com/grizli-v2/JwstMosaics/v7/index.html}. The CEERS catalog includes 7 {\it HST} bands  (F435W, F606W, F814W, F105W, F125W, F140W,
and F160W), and 7 {\it JWST}/NIRCam bands
(F115W, F150W, F200W, F277W, F356W, F410M,
and F444W), while the JADES catalog includes 9 {\it HST} bands  (F435W, F606W, F775W F814W, F850LP, F105W, F125W, F140W,
and F160W), and 14 {\it JWST}/NIRCam bands
(both the broadband filters F090W, F115W, F150W, F200W, F277W, F356W, and F444W, and the medium-band filters F335M and F410M, supplemented by F180M, F210M, and F444W photometry from FRESCO \citep{2023MNRAS.525.2864O,https://doi.org/10.17909/gdyc-7g80}. We additionally included medium-band imaging from the JEMS program \citep{https://doi.org/10.17909/fsc4-dt61,2023ApJS..268...64W} in the F182M, F210M, F430M, F460M and F480M filters.) For objects lacking NIRCam coverage, 
we used the SEDs cataloged by the 3D-HST team \citep{2014ApJS..214...24S,2016ApJS..225...27M} 
in the AEGIS and GOODS-S fields for CEERS and JADES, respectively. These SEDs include ground-based and {\it HST} optical and near-IR photometry, and {\it Spitzer}/IRAC $3.6-8.0$~$\mu$m measurements.

The SED fitting was done using the FAST \citep{Kriek_2009} algorithm with the FSPS models from \citet{2009ApJ...699..486C} assuming a delayed-$\tau$ star-formation history. Following the redshift and stellar-mass-dependent criteria described in Section 2.2 of \citet{2018ApJ...860...75D}, we assume either a 1.4-times solar metallicity and a \citet{2000ApJ...533..682C} dust extinction law (hereafter $1.4Z_{\odot}$+Calz), or a sub-solar metallicity and a Small Magellanic Cloud \citep[SMC;][]{2003ApJ...594..279G} extinction curve (hereafter 0.28$Z_{\odot}$+SMC). For galaxies in the range $1.4 < z \leq 2.7$, we assumed 0.28$Z_{\odot}$+SMC for galaxies below $10^{10.45}\ M_{\odot}$, and $1.4Z_{\odot}$+Calz otherwise. For galaxies in the range $2.7 < z \leq 3.4$, we assumed 0.28$Z_{\odot}$+SMC for galaxies below $10^{10.66}\ M_{\odot}$, and $1.4Z_{\odot}$+Calz otherwise. For all galaxies above $z=3.4$, we assumed 0.28$Z_{\odot}$+SMC. The best-fit SED was then used to model the continuum for each galaxy during the initial line fitting procedure. Using the line fluxes calculated during this initial line fitting procedure, we corrected the photometry for contributions from emission lines that were detected with $>$5$\sigma$ significance. The SEDs were then re-fit using FAST on the emission-line-corrected photometry. The resulting best-fit SEDs were used to model the continuum during the final run of the line fitting.

The emission lines were fit using a custom Python procedure that utilizes \texttt{scipy.optimize.curve\_fit()} to perform a chi-squared minimization with a Levenberg-Marquardt algorithm. Each emission line was modeled individually with a single Gaussian profile; however, adjacent emission lines (e.g., H$\gamma$ and \oiii $\lambda 4364$, or H$\alpha$ and \nii $\lambda\lambda 6550,6585$) were fit simultaneously. The flux ratio of the \nii$\lambda\lambda 6550,6585$ line doublet was fixed with a ratio of 1:3 during the fitting procedure since the intrinsic line strengths of the doublet members are fixed to the ratio of their Einstein A coefficients. The 68\% confidence interval (and associated 1$\sigma$ uncertainties) of the line fluxes were obtained via Monte Carlo simulations, perturbing the flux density spectrum according to the error spectrum 500 times and calculating the 16th and 84th percentiles of the resulting flux distribution.

\begin{figure}[h!]
    \centering
    \includegraphics[width=8.5cm]{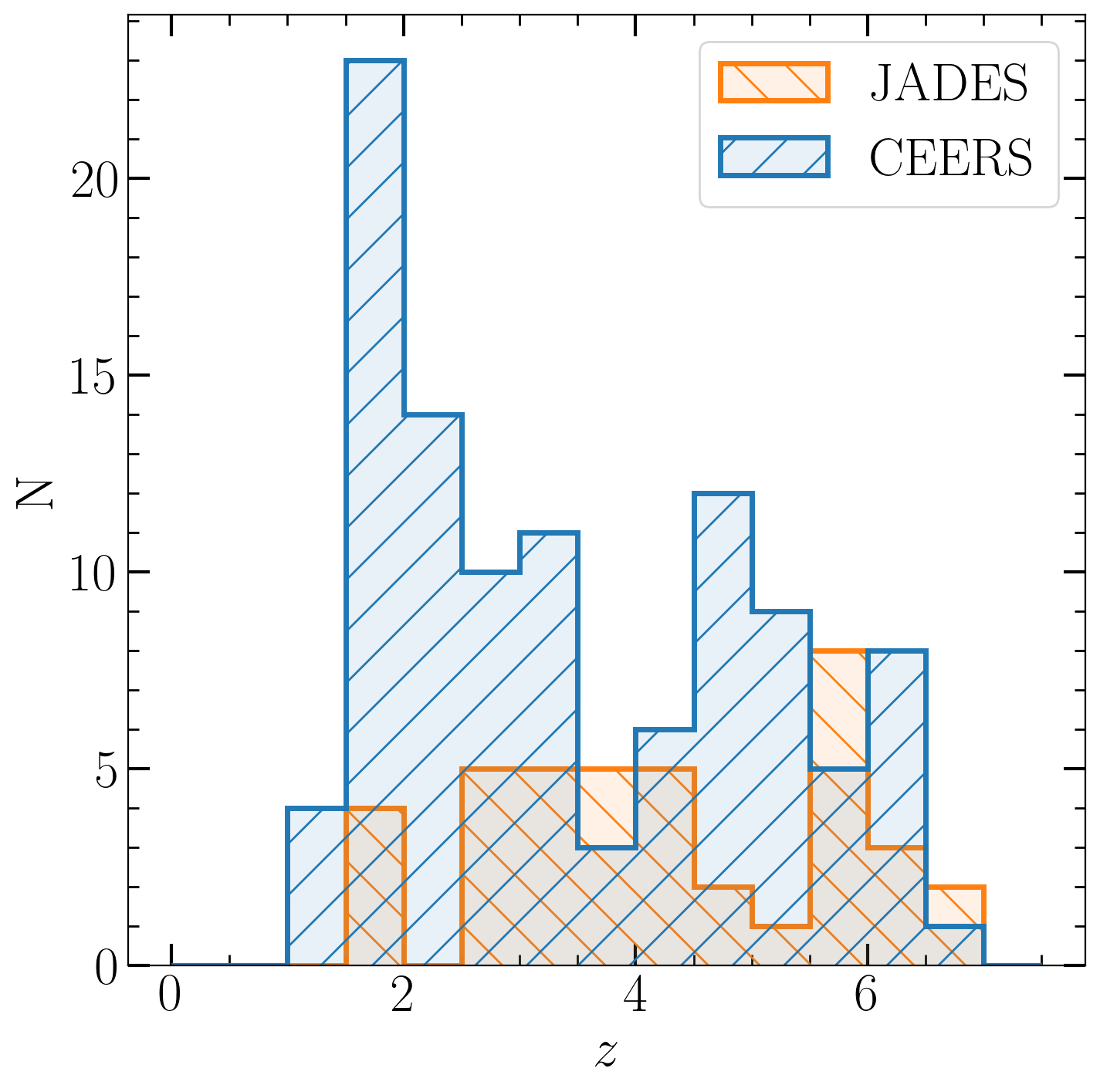}
    \caption{Redshift distribution of the combined CEERS and JADES sample that we analyze consisting of 146 spectroscopically confirmed star-forming galaxies with robust SED measurements. Of these, 106 galaxies come from CEERS and 40 galaxies come from JADES.}
    \label{fig:zdist}
\end{figure}

\subsection{Sample Properties and SFR calculations}

Of the 117 JADES galaxies with measured redshifts, in addition to ground-based photometry, 110 have JWST/NIRCam and 3DHST photometry, 3 have only 3DHST photometry, and 4 have no space-based photometric observations. Of the 231 galaxies with measured redshifts in the CEERS sample, 217 had a combination of NIRCam and/or 3DHST measurements. We further restricted our sample to galaxies with $>$5$\sigma$ detections of at least two Balmer lines for dust correction, and we selected for star-forming galaxies by imposing a cutoff in specific SFR of $\log{\rm(sSFR(SED)/yr^{-1})} > -11$, where sSFR(SED) is the sSFR calculated from the SED fitting routine. Finally, we flagged and removed AGN-dominated galaxies by identifying either an \nii/H$\alpha$ flux ratio $>$0.5, or through the presence of broad Balmer emission. Overall, the combined CEERS and JADES sample that we analyze consists of 146 galaxies in the redshift range $1.4\leq z < 7$.

\begin{figure*}[ht]
    \centering
    \includegraphics[width=17cm]{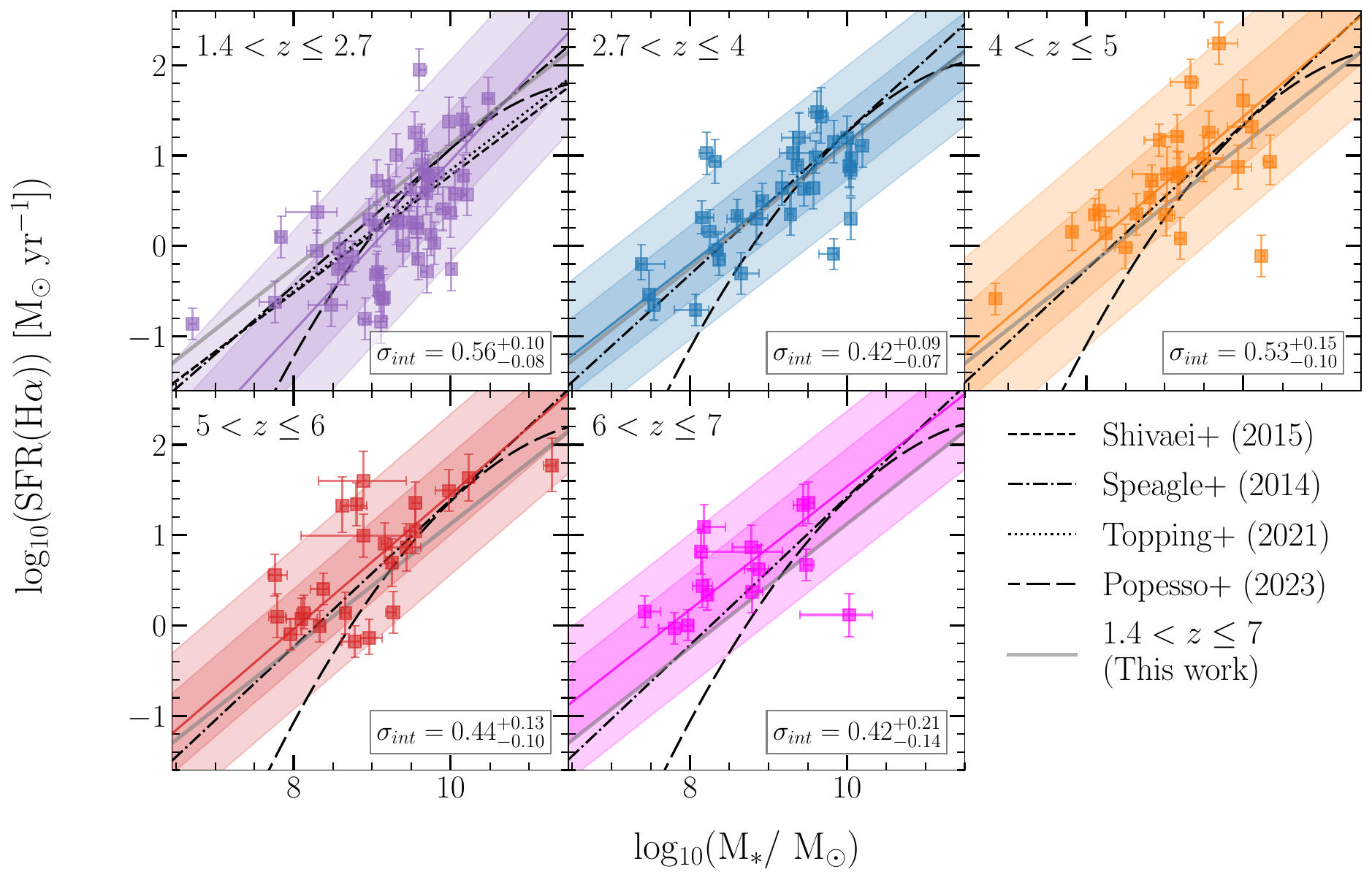}
    \caption{H$\alpha$-based SFRs vs. stellar masses for the combined JADES and CEERS sample binned by redshift. The darker and lighter shaded regions denote the 1$\sigma_{\rm int}$ and 2$\sigma_{\rm int}$ intrinsic scatter intervals about the SFMS, respectively. The thick gray line shows the best-fit SFMS for the whole sample in the range $1.4 < z \leq 7$. The literature curves from \citet{2015ApJ...815...98S}, \citet{2014ApJS..214...15S}, \citet{2021MNRAS.506.1237T}, and \citet{2023MNRAS.519.1526P} were shifted down in SFR by 0.32, 0.32, 0.37, and 0.32 dex, respectively to match our low-metallicity H$\alpha$ to SFR conversion. The slightly different shift for \citet{2021MNRAS.506.1237T} reflects the use of the \citet{2011ApJ...741..124H} conversion, whereas the other works adopt the conversion from \citet{1998ARA&A..36..189K}.}
    \label{fig:msha}
\end{figure*}

\begin{figure*}
    \centering
    \includegraphics[width=17cm]{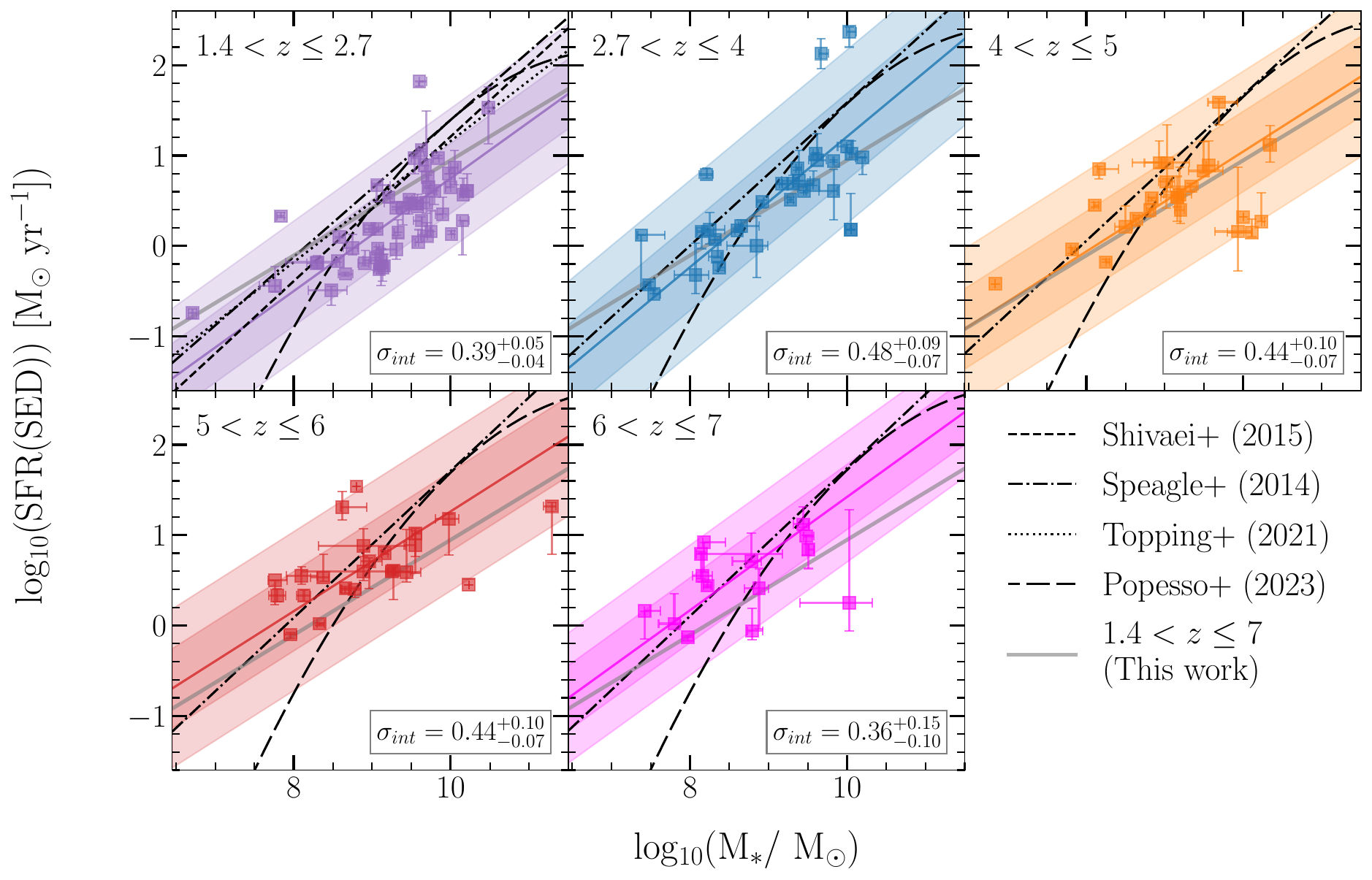}
    \caption{SED-based SFRs vs. stellar masses for the combined JADES and CEERS sample binned by redshift. The darker and lighter shaded regions denote the 1$\sigma_{\rm int}$ and 2$\sigma_{\rm int}$ intrinsic scatter intervals about the SFMS, respectively. The thick gray line shows the best-fit SFMS for the whole sample in the range $1.4 < z \leq 7$. The lower normalization of our sample compared to the literature is largely due to the fact that we assume a SMC dust law, whereas most literature studies have assumed a Calzetti dust law.}
    \label{fig:mssed}
\end{figure*}

\begin{figure*}
    \centering
    \includegraphics[width=17cm]{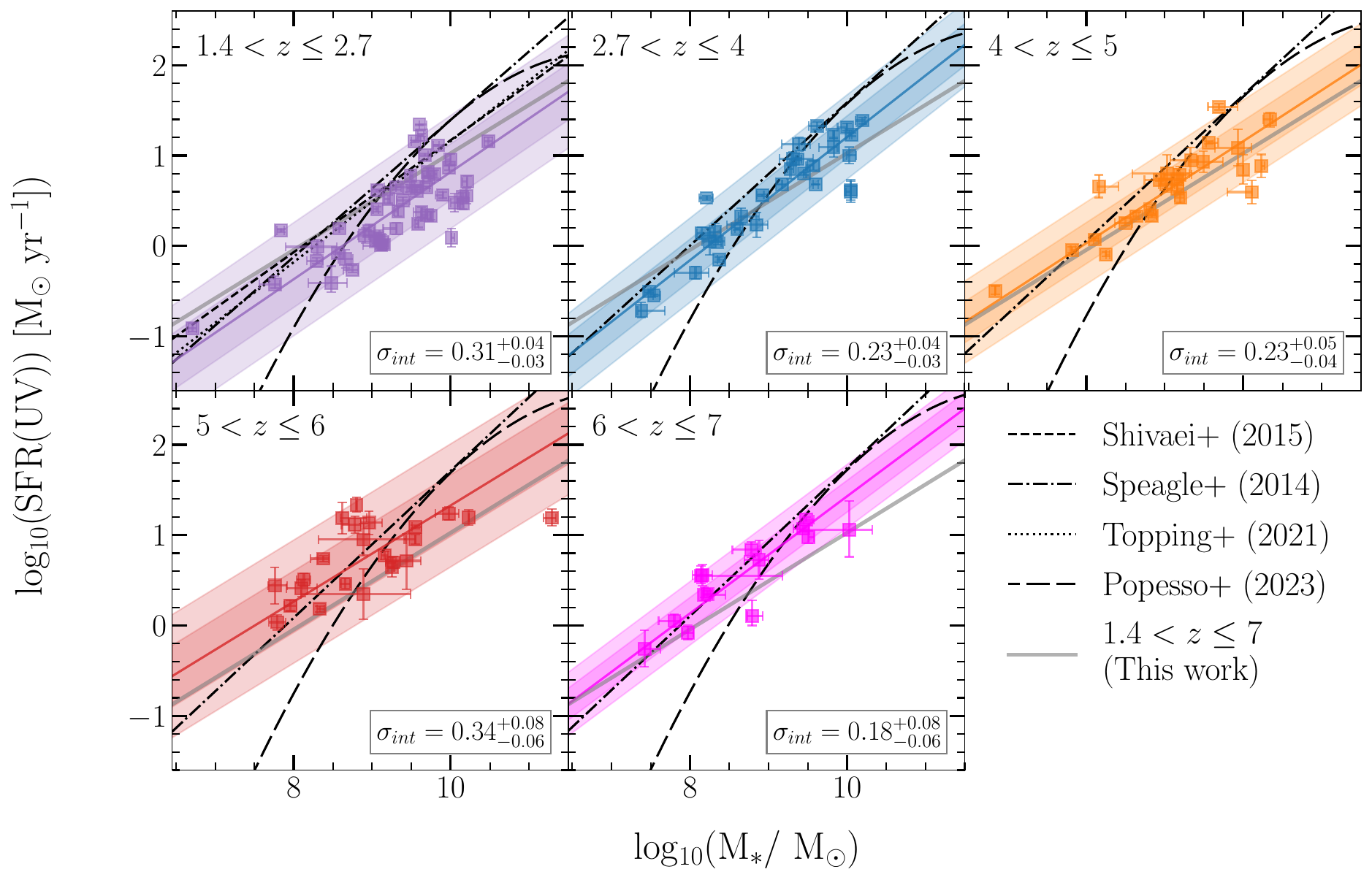}
    \caption{UV-based SFRs vs. stellar masses for the combined JADES and CEERS sample binned by redshift. The darker and lighter shaded regions denote the 1$\sigma_{\rm int}$ and 2$\sigma_{\rm int}$ intrinsic scatter intervals about the SFMS, respectively. The thick gray line shows the best-fit SFMS for the whole sample in the range $1.4 < z \leq 7$. The lower normalization of our sample compared to the literature is largely due to the fact that we assume a SMC dust law, whereas most literature studies have assumed a Calzetti dust law.}
    \label{fig:msuv}
\end{figure*}

We show the redshift distribution of the combined JADES and CEERS NIRSpec sample in Figure \ref{fig:zdist}. The full sample extends between $z\sim 1-7$, providing insights into the high-redshift universe enabled by the near-IR sensitivity that JWST/NIRSpec provides. Prior to {\it JWST}, the properties and evolution of star-forming galaxies from the local universe out to $z\sim 3$ had been investigated extensively in the literature. Our sample, however, contains 94 galaxies at $z>2.7$ with $>$5$\sigma$ Balmer line detections, Balmer-decrement-based E($B-V$) measurements, and SED information, enabling an analysis of the properties of star-forming galaxies both during and after the epoch of reionization. We calculated the SFRs of the galaxies in our sample using the dust-corrected H$\alpha$ luminosity, the SED-based SFR from FAST, and the dust-corrected UV luminosity. To correct the observed H$\alpha$ luminosity for dust attenuation, we assumed a \citet{1989ApJ...345..245C} dust law, with E($B-V$) values derived from the H$\alpha$/H$\beta$ Balmer decrement. We assumed an intrinsic ratio of H$\alpha$/H$\beta$ of 2.79, corresponding to Case B recombination where $T_e = 15,000\ \rm K$. In cases where H$\alpha$ was not detected due to lack of wavelength coverage (15 galaxies), we used the H$\gamma$/H$\beta$ ratio to dust-correct the emission lines, assuming an intrinsic ratio of 0.475. Then, using the dust-corrected H$\beta$ flux, we inferred the H$\alpha$ flux with the assumed intrinsic H$\alpha$/H$\beta$ ratio of 2.79. In cases where H$\beta$ was not detected due to lack of wavelength coverage (13 galaxies), we used the H$\gamma$/H$\alpha$ ratio to calculate E($B-V$), assuming an intrinsic ratio of 0.173. All Balmer line fluxes have been measured accounting for Balmer absorption in the underlying stellar continuum during the fitting procedure.

After correcting for dust attenuation, we converted the H$\alpha$ luminosities to SFRs according to the following equation:

\begin{equation}
    \log\left({\rm {\frac{SFR}{M_\odot\ yr^{-1}}}} \right) = \log\left({\rm \frac{L_{H\alpha}}{erg\ s^{-1}}} \right) + C
\end{equation}

where $C$ is a metallicity-dependent conversion calculated from a set of BPASS models \citep{2018MNRAS.479...75S} with an upper stellar mass limit of 100 M$_\odot$ \citep{2022ApJ...926...31R}. For galaxies fit with the $1.4Z_{\odot}$+Calz assumption, we used a conversion of $C=-41.37$, corresponding to a stellar metallicity of $Z_*=0.02$. For galaxies fit with the 0.28$Z_{\odot}$+SMC assumption, we used a conversion of $C=-41.59$, corresponding to a stellar metallicity of $Z_*=0.004$. To calculate a final uncertainty on the measured H$\alpha$-based SFR, we add the corresponding absolute flux calibration uncertainty in quadrature with the measurement uncertainties, adding 0.17 dex for those galaxies in JADES, and 0.23 dex for those in CEERS. 

In order to correct the UV luminosity for dust attenuation, we calculated the UV slope, $\beta$, by performing a linear regression to the observed rest-UV photometry for each galaxy, which yields a median beta value of $-1.55$ with a median uncertainty on beta of 0.36. We converted $\beta$ to the attenuation at 1600 \AA, $A_{1600}$, using the following metallicity- and dust-law-dependent conversion from \citet{2018ApJ...853...56R}:

\begin{equation}
    A_{1600} = 
    \begin{cases}
        1.82 \beta + 4.43, & {\rm 1.4Z_{\odot}}\text{+Calz} \\
        0.93 \beta + 2.52, & 0.28 {\rm Z_{\odot}}\text{+SMC}
    \end{cases}
\end{equation}

We then converted the dust-corrected UV luminosity to a SFR using the \citet{2011ApJ...737...67M} and \citet{2011ApJ...741..124H} conversion:

\begin{equation}\label{eq:sfr_uv}
    \log\left( {\rm \frac{SFR}{M_\odot\ yr^{-1}}} \right) = \log\left( {\rm \frac{\nu L_{\nu, 1600}}{erg\ s^{-1}}} \right) - 43.35
\end{equation}

adjusted for a \citet{2003PASP..115..763C} IMF by adding $\log_{10}(0.77)$ to equation \ref{eq:sfr_uv}. We plot the SFRs calculated using H$\alpha$, SED-fitting, and UV luminosity vs. the stellar masses in Figures \ref{fig:msha}, \ref{fig:mssed}, and \ref{fig:msuv}, respectively.

In our analysis of the SFMS, we divide the galaxy sample into bins of redshift in order to track changes in galaxy properties with cosmic time. We restrict our analysis to objects with $z\geq 1.4$ for the purpose of comparison to previous related work from the MOSDEF survey \citep[e.g.][]{2015ApJ...815...98S,2021MNRAS.506.1237T}. We define our lowest redshift bin, $1.4<z\leq 2.7$, to capture galaxies up to $z=2.7$, beyond which the H$\alpha$ line is shifted out of the wavelength range of ground-based observatories. The following redshift bins ($2.7<z\leq 4$, $4<z\leq 5$, $5<z\leq 6$, and $6<z\leq 7$) represent the new rest-optical redshift frontier enabled by JWST/NIRSpec.

\section{Fitting The Star-forming Main Sequence}

We performed a linear fit to the SFMS in Figures \ref{fig:msha}, \ref{fig:mssed}, and \ref{fig:msuv} assuming an equation of the form:

\begin{align}\label{eq:sfms}
    \nonumber \log({\rm SFR}) &= m\times \log \left({\rm \frac{M_*}{10^{9.16} M_{\odot}}} \right) \\
    &  + b + \mathcal{N}(0, \rm \sigma_{int}^2) 
\end{align}

where $m$ is the slope, $b$ is the intercept, $10^{9.16}$ is the median stellar mass of the sample, and $\rm \sigma_{int}$ is the intrinsic scatter distributed normally ($\mathcal{N}$) about the best-fit relation. We normalized the stellar masses by $10^{9.16}$, the median stellar mass, during the fitting to minimize the covariance between the best-fit slope and the intercept. Previous works have found evidence for a turnover in the SFMS at high masses, leading some authors to use a different parametrization \citep[e.g.,][]{2015ApJ...801...80L,2023MNRAS.519.1526P}. However, our galaxy sample predominantly lies below the characteristic mass of $\sim$$10^{10.5}\ M_\odot$ where this turnover occurs, so we proceed using a linear fit.

We estimated the parameters $m$, $b$, and $\rm \sigma_{int}$ using the following likelihood function $\mathcal{L}$ presented by \citet{2010arXiv1008.4686H}:

\begin{align}\label{eq:likelihood}
    \ln{\mathcal{L}} &= - \sum_{i=1}^N \frac{1}{2}\ln[\Sigma_i^2 + V]  - \sum_{i=1}^N \frac{\Delta_i^2}{2[\Sigma_i^2+V]}\\
    \Delta_i  &= \boldsymbol{\hat{\upsilon}}^T 
    \begin{bmatrix}
        x_i\\
        y_i
    \end{bmatrix}
    - \frac{b}{\sqrt{1+m^2}}\\
    \boldsymbol{\hat{\upsilon}} &= \frac{1}{\sqrt{1+m^2}}
    \begin{bmatrix}
        -m\\
        1
    \end{bmatrix}
\end{align}

% \begin{align}
    
% \end{align}

where $x_i = \log{(\rm M_*/10^{9.16}\ M_\odot)}_i$, $y_i=\log{\rm (SFR/M_\odot\ yr^{-1})}_i$, $\Delta_i$ is the orthogonal distance of each point $(x_i, y_i)$ to the best-fit line, $N$ is the number of data points, $\boldsymbol{\hat{\upsilon}}$ is the unit vector orthogonal to the best-fit relation, $\Sigma_i^2$ is the covariance matrix of each measurement projected onto the best-fit line, defined as:

\begin{align}
    \Sigma_i^2 = \boldsymbol{\hat{\upsilon}}^T
    \begin{bmatrix}
        \sigma_{x,i}^2 & \sigma_{xy,i} \\
        \sigma_{xy,i} & \sigma_{y,i}^2
    \end{bmatrix}
    \boldsymbol{\hat{\upsilon}}
\end{align}

with $\sigma_{x,i}$ being the uncertainty on $x_i$, $\sigma_{y,i}$ being the uncertainty on $y_i$, $\sigma_{xy,i}$ being the covariance between the two measurements for a given galaxy, and $V$ is a matrix encapsulating the intrinsic scatter, defined as:

\begin{equation}
    V = \boldsymbol{\hat{\upsilon}}^T
    \begin{bmatrix}
        0 & 0 \\
        0 & \rm \sigma_{int}^2
    \end{bmatrix}
    \boldsymbol{\hat{\upsilon}}
\end{equation}

Throughout this analysis, we assume that $\sigma_{xy,i} = 0$ for all galaxies. We emphasize that this is likely not the case when the SFR is estimated using SED fitting, since these two quantities are not derived independently. However, we do not quantify this covariance, and instead we proceed with this caveat in mind when analyzing the SED-based SFMS.

We multiply our likelihood function $\mathcal{L}$ by a uniform prior probability distribution, restricting each parameter to the following ranges: $m=[0.5,1.5]$, $b=[-5,5]$, and $\sigma_{\rm int}=[0,1.5]$. To estimate the best-fit parameters, we use the Python Markov Chain Monte Carlo (MCMC) implementation \texttt{emcee}\footnote{https://emcee.readthedocs.io/en/stable/} developed by \citet{2013PASP..125..306F}. We use 48 walkers and a maximum of 5000 steps to sample the likelihood function, always discarding the first 100 ``burn-in" samples before evaluating the best-fit parameters. These hyperparameters ensure that the number of steps is always greater than 50 times the autocorrelation time. We evaluate the likelihood function within the bounds defined by the priors, and we calculate the best-fit parameters for the SFMS in each redshift bin for each SFR indicator. For each parameter, we report the median value of the MCMC sample distribution, and we calculate the $1\sigma$ confidence interval by reporting the 16th and the 84th percentile values. The results of this analysis are given in Table \ref{tab:fitparams}.

\begin{figure}
    \centering
    \includegraphics[width=8.5cm]{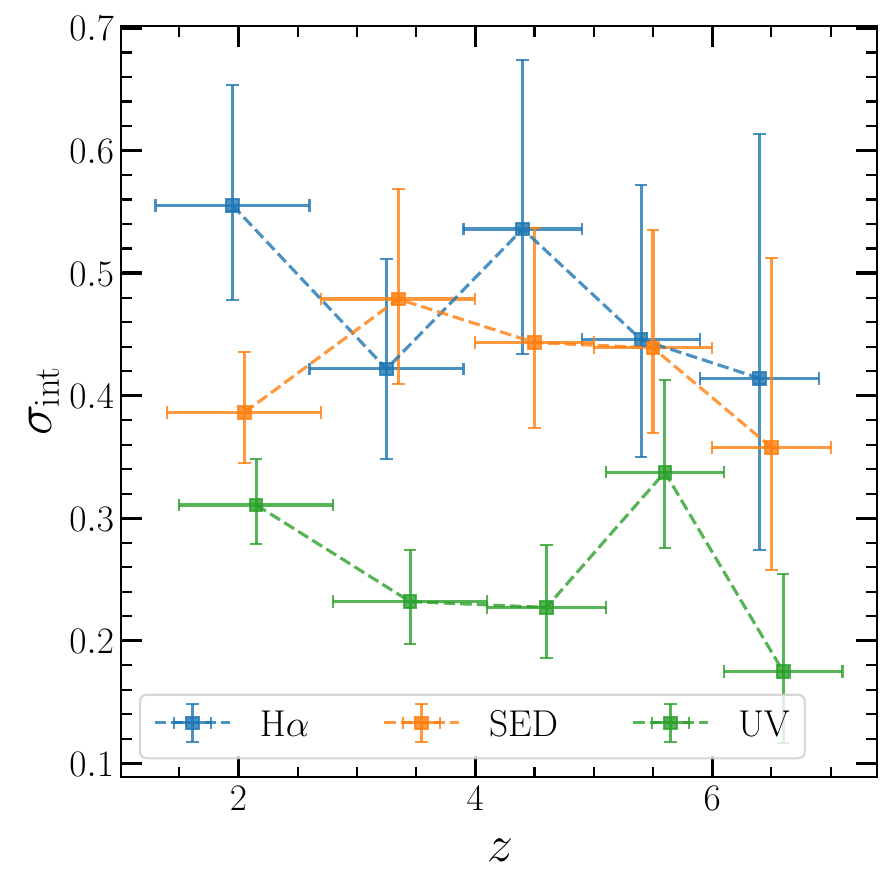}
    \caption{The intrinsic scatter in the SFMS $\rm \sigma_{int}$ as a function of redshift. The points have an arbitrary offset in the x-axis to more easily see overlapping points. The blue, orange, and green points show the intrinsic scatter in the SFMS as determined from H$\alpha$, SED fitting, and UV luminosity, respectively.}
    \label{fig:sig_int_z}
\end{figure}

For each SFR indicator, we find a consistent level of intrinsic scatter at all redshifts, with the inverse-variance-weighted average values being $\langle \sigma_{{\rm int,H}\alpha} \rangle = 0.48 \pm 0.05$, $\langle \sigma_{{\rm int,SED}} \rangle = 0.42\pm 0.03$, and $\langle \sigma_{{\rm int,UV}} \rangle = 0.26\pm 0.02$. In Figure \ref{fig:sig_int_z}, we illustrate how our $\rm \sigma_{int}$ measurements vary over each redshift bin for each SFR indicator. We see that over the majority of the redshift range that we probe, $\sigma_{\rm int, H\alpha}$ and $\sigma_{\rm int, SED}$ are greater than $\sigma_{\rm int, UV}$.

The scatter in the SED-based SFMS is more consistent with the H$\alpha$-based SFMS than with the UV-based SFMS, contrary to our expectations given that the SED-based and UV-based SFRs are both measured using stellar continuum emission. However, the SED-based SFMS scatter may be overestimated due to anticorrelated errors between the stellar mass and SFR in the SED fitting. Since the SFR and stellar mass are both sensitive to the normalization of the SED model to the photometry, the two are highly correlated. If there is positive covariance between the stellar mass and the SFR uncertainties from SED fitting, then setting $\rm \sigma_{xy,i} = 0$ and minimizing Equation \ref{eq:likelihood} will underestimate the intrinsic scatter. However, some studies \citep[e.g.,][]{2015ApJ...799..183S,2016ApJ...820L...1K} have found in the analysis of their SED fitting approach that the SFR vs. stellar mass covariance was negative, driven largely by the uncertainties on the UV extinction. This negative covariance would lead to an overestimate of the intrinsic scatter in the SED-based SFMS using our methodology. Since we do not explicitly test for the covariance between the stellar mass and the SED-based SFR, it is difficult to assess the accuracy of our $\rm \sigma_{int,SED}$ calculations. However, since the measurements of SFR(H$\alpha$) and SFR(UV) are made independently of the SED fitting process, we proceed assuming that the uncertainties in the H$\alpha$- and UV-based SFRs are not correlated, or are only weakly correlated with the stellar mass uncertainties.

In each redshift bin, we additionally calculate the intrinsic scatter below the median stellar mass of the sample ($\rm M_* < 10^{9.16}\ M_{\odot}$) and above the median stellar mass of the sample ($\rm M_* > 10^{9.16}\ M_{\odot}$) to investigate whether we detect the mass-dependence of $\rm \sigma_{int}$ reported by other studies \citep[e.g.,][]{2023arXiv231210152C,2017ApJ...847...76S}. To calculate the mass-dependent $\rm \sigma_{int}$, we calculate the maximum likelihood value of Equation \ref{eq:likelihood}, treating $V$ as a piecewise function:

\begin{equation}
    V = 
    \begin{cases}
        V({\rm \sigma_{int,low}}), & {\rm M_* < 10^{9.16}\ M_{\odot}}\\
        V({\rm \sigma_{int,high}}), & {\rm M_* > 10^{9.16}\ M_{\odot}}
    \end{cases}
\end{equation}

We then performed 500 Monte Carlo simulations, perturbing each data point $(x_i, y_i)$ by the measurement uncertainty $(\sigma_{x,i}, \sigma_{y,i})$ and allowing galaxies to move between mass bins. The results of our simulations are also tabulated in Table \ref{tab:fitparams}. We do not find strong evidence for a mass dependence of the intrinsic scatter in the SFMS based on this analysis.

\begin{deluxetable*}{cc|ccccccc}
\label{tab:fitparams}
\caption{Star-forming Main Sequence best-fit parameters as a function of redshift and SFR indicator.}
\tablehead{
\colhead{SFR Indicator} & \colhead{Redshfit Bin} & \colhead{m} & \colhead{b} & \colhead{$\sigma_{\rm int}$} & \colhead{$\sigma_{\rm int,low}$$\tablenotemark{\scriptsize a}$} & \colhead{$\sigma_{\rm int,high}$$\tablenotemark{\scriptsize b}$} & \colhead{N}
}
    \startdata
    H$\alpha$ & $1.4 < z \leq 7.0$ & $0.69^{+0.07}_{-0.08}$ & $0.56^{+0.05}_{-0.05}$ & $0.53^{+0.05}_{-0.04}$ & $0.57^{+0.05}_{-0.04}$ & $0.59^{+0.04}_{-0.04}$ & 146\\
H$\alpha$ & $1.4 < z \leq 2.7$ & $0.95^{+0.15}_{-0.18}$ & $0.16^{+0.09}_{-0.09}$ & $0.56^{+0.10}_{-0.08}$ & $0.89^{+0.33}_{-0.28}$ & $0.58^{+0.06}_{-0.06}$ & 52\\
H$\alpha$ & $2.7 < z \leq 4.0$ & $0.67^{+0.10}_{-0.12}$ & $0.59^{+0.09}_{-0.08}$ & $0.42^{+0.09}_{-0.07}$ & $0.45^{+0.07}_{-0.08}$ & $0.44^{+0.07}_{-0.07}$ & 32\\
H$\alpha$ & $4.0 < z \leq 5.0$ & $0.74^{+0.15}_{-0.20}$ & $0.81^{+0.12}_{-0.12}$ & $0.54^{+0.15}_{-0.10}$ & $0.26^{+0.09}_{-0.10}$ & $0.74^{+0.11}_{-0.09}$ & 25\\
H$\alpha$ & $5.0 < z \leq 6.0$ & $0.75^{+0.13}_{-0.16}$ & $0.83^{+0.12}_{-0.11}$ & $0.44^{+0.13}_{-0.10}$ & $0.51^{+0.08}_{-0.08}$ & $0.32^{+0.12}_{-0.13}$ & 23\\
H$\alpha$ & $6.0 < z \leq 7.0$ & $0.69^{+0.13}_{-0.26}$ & $0.97^{+0.20}_{-0.15}$ & $0.42^{+0.21}_{-0.14}$ & $0.26^{+0.10}_{-0.11}$ & $0.62^{+0.21}_{-0.21}$ & 14\\
\hline
SED & $1.4 < z \leq 7.0$ & $0.53^{+0.02}_{-0.04}$ & $0.51^{+0.04}_{-0.04}$ & $0.45^{+0.03}_{-0.03}$ & $0.44^{+0.02}_{-0.02}$ & $0.47^{+0.02}_{-0.02}$ & 146\\
SED & $1.4 < z \leq 2.7$ & $0.62^{+0.07}_{-0.09}$ & $0.23^{+0.06}_{-0.06}$ & $0.39^{+0.05}_{-0.04}$ & $0.34^{+0.03}_{-0.03}$ & $0.39^{+0.02}_{-0.01}$ & 52\\
SED & $2.7 < z \leq 4.0$ & $0.72^{+0.11}_{-0.14}$ & $0.61^{+0.09}_{-0.09}$ & $0.48^{+0.09}_{-0.07}$ & $0.29^{+0.03}_{-0.03}$ & $0.54^{+0.04}_{-0.04}$ & 32\\
SED & $4.0 < z \leq 5.0$ & $0.55^{+0.04}_{-0.08}$ & $0.59^{+0.10}_{-0.09}$ & $0.44^{+0.10}_{-0.07}$ & $0.27^{+0.03}_{-0.03}$ & $0.55^{+0.07}_{-0.05}$ & 25\\
SED & $5.0 < z \leq 6.0$ & $0.55^{+0.04}_{-0.09}$ & $0.81^{+0.10}_{-0.10}$ & $0.44^{+0.10}_{-0.07}$ & $0.40^{+0.04}_{-0.03}$ & $0.41^{+0.06}_{-0.05}$ & 23\\
SED & $6.0 < z \leq 7.0$ & $0.63^{+0.09}_{-0.17}$ & $0.90^{+0.14}_{-0.12}$ & $0.36^{+0.15}_{-0.10}$ & $0.36^{+0.07}_{-0.06}$ & $0.08^{+0.11}_{-0.08}$ & 14\\
\hline
UV & $1.4 < z \leq 7.0$ & $0.54^{+0.03}_{-0.04}$ & $0.58^{+0.03}_{-0.03}$ & $0.34^{+0.02}_{-0.02}$ & $0.35^{+0.02}_{-0.01}$ & $0.34^{+0.01}_{-0.01}$ & 145\\
UV & $1.4 < z \leq 2.7$ & $0.59^{+0.05}_{-0.07}$ & $0.32^{+0.04}_{-0.05}$ & $0.31^{+0.04}_{-0.03}$ & $0.24^{+0.02}_{-0.02}$ & $0.33^{+0.01}_{-0.01}$ & 52\\
UV & $2.7 < z \leq 4.0$ & $0.68^{+0.05}_{-0.06}$ & $0.63^{+0.05}_{-0.05}$ & $0.23^{+0.04}_{-0.03}$ & $0.20^{+0.02}_{-0.02}$ & $0.26^{+0.03}_{-0.03}$ & 31\\
UV & $4.0 < z \leq 5.0$ & $0.56^{+0.04}_{-0.06}$ & $0.68^{+0.05}_{-0.05}$ & $0.23^{+0.05}_{-0.04}$ & $0.16^{+0.03}_{-0.02}$ & $0.30^{+0.04}_{-0.04}$ & 25\\
UV & $5.0 < z \leq 6.0$ & $0.53^{+0.03}_{-0.05}$ & $0.88^{+0.08}_{-0.07}$ & $0.34^{+0.07}_{-0.06}$ & $0.32^{+0.06}_{-0.04}$ & $0.30^{+0.07}_{-0.07}$ & 23\\
UV & $6.0 < z \leq 7.0$ & $0.64^{+0.08}_{-0.09}$ & $0.88^{+0.08}_{-0.07}$ & $0.18^{+0.08}_{-0.06}$ & $0.22^{+0.05}_{-0.06}$ & $0.08^{+0.04}_{-0.08}$ & 14\\
    \enddata
\tablenotetext{*}{We are unable to measure a UV-based SFR for one galaxy in the $2.7 < z \leq 4.0$ bin due to a lack of rest-UV photometric observations.}
\tablenotetext{a}{The intrinsic scatter in the SFMS as measured for galaxies with stellar masses less than the sample median of $10^{9.16}\ M_\odot$.}
\tablenotetext{b}{The intrinsic scatter in the SFMS as measured for galaxies with stellar masses greater than the sample median of $10^{9.16}\ M_\odot$.}
\end{deluxetable*}

\section{Discussion} \label{sec:discussion}

\subsection{SFR Intrinsic Scatter}
We analyze both H$\alpha$- and UV-based SFRs in order to characterize the burstiness of galaxy SFHs in the early universe. We see in Figure \ref{fig:sig_int_z} that the H$\alpha$-based SFMS is characterized by a larger intrinsic scatter than the UV-based SFMS, with an average difference of $\langle \sigma_{{\rm int,H}\alpha} \rangle - \langle \sigma_{{\rm int,UV}} \rangle = 0.21 \pm 0.05$ dex. We argue that this difference in $\rm \sigma_{int}$ is consistent with time fluctuations in galaxy SFHs, though systematic uncertainties may also contribute (see Section \ref{sec:confounding_factors}).

\subsubsection{Comparison to theoretical works}
Some previous theoretical works have predicted that galaxies should have bursty SFHs at high redshift. \citet{2015MNRAS.451..839D} analyzed hydrodynamic simulations \citep[{\sc gasoline};][]{2004NewA....9..137W} of 11 star-forming galaxies and calculated the standard deviation in the H$\alpha$ luminosity values in 5 Myr time steps from $z=2.74$ to $z=1.97$. They found that the standard deviation in the H$\alpha$ luminosities ($L_{\rm H\alpha}$) of galaxies between $10^8$ and $10^9\ M_\odot$ was 0.45 dex, while that for galaxies with stellar masses $\gtrsim$$10^9\ M_\odot$ was 0.22 dex. We compare these predictions to our observations of the H$\alpha$-based SFMS in the $1.4<z\leq 2.7$ bin assuming that our measurements of $\sigma_{\rm int,H\alpha}$ (and thus, $L_{\rm H\alpha}$) capture the average deviations in SFR from a shared SFH. The scatter in $L_{\rm H\alpha}$ that \citet{2015MNRAS.451..839D} found in their $10^8\text{-}10^9\ M_\odot$ galaxies (0.45 dex) is lower than our SFR(H$\alpha$) intrinsic scatter of $0.56^{+0.10}_{-0.08}$ dex measured across the full mass range of our sample. This discrepancy also holds with respect to $\sigma_{\rm int, H\alpha}$ in our lower-mass bin ($<$$10^{9.16}\ M_\odot$), which we determine to be $0.90^{+0.32}_{-0.28}$ dex. However, the uncertainty on $\sigma_{\rm int, H\alpha}$ in the lower mass bin has large uncertainties owing to the relatively few galaxies in this mass range and the measurement uncertainties moving galaxies between mass bins. With respect to the higher-mass ($\gtrsim$$10^9\ M_\odot$) sample of \citet{2015MNRAS.451..839D}, their predicted standard deviation in $L_{H\alpha}$ of 0.22 dex is lower than what we find for $\sigma_{\rm int,H\alpha}$ across the whole mass range in our sample as well as in our high stellar mass bin ($0.56^{+0.10}_{-0.08}$ and $0.58^{+0.06}_{-0.06}$, respectively). 

When considering the UV-based SFMS, the standard deviation in $L_{UV}$ that \citet{2015MNRAS.451..839D} predicted also presents some differences when compared with our measurements. The predicted SFR(UV) scatter from \citeauthor{2015MNRAS.451..839D} was 0.28 dex at $M_*=10^8\text{--}10^9\ M_\odot$, and 0.16 at $M_*\gtrsim$10$^9\ M_\odot$. The scatter in their $10^8\text{--}10^9\ M_\odot$ sample is consistent with our measurement of $\sigma_{\rm int, UV} = 0.31^{+0.04}_{-0.03}$ across our full mass range within the 1$\sigma$ confidence interval. However, the scatter in our low-mass bin of $0.24^{+0.02}_{-0.02}$ dex is slightly smaller than the 0.28 dex prediction. Whether this difference comes down to small sample numbers in our low-mass bin or from differences in galaxy SFHs is still uncertain, and more observations of galaxies at lower masses will improve this scatter measurement. In their $\gtrsim$10$^9\ M_\odot$ galaxies, the predicted scatter of 0.16 is smaller than what we find in both our high-mass galaxies ($0.33^{+0.01}_{-0.01}$ dex) and across the whole mass range. 

Our finding that the scatter in $L_{\rm H\alpha}$ predicted by \citet{2015MNRAS.451..839D} is smaller than our measured $\sigma_{\rm int,H\alpha}$ suggests that systematic measurement uncertainties may be contributing to our inferred scatter, or that the output from these simulations underpredicts the stochasticity of SFRs in galaxies between $z=2.74$ and $z=1.97$. A larger, mass-complete spectroscopic sample would help to distinguish between these interpretations.

We also compare our data to the intrinsic scatter values measured in the FIRE and IllustrisTNG simulations. In the FIRE simulations, \citet{2017MNRAS.466...88S} reports the predicted scatter in the SFMS to be between 0.4--0.5 dex for H$\alpha$-based SFRs and 0.3--0.4 dex for UV-based SFRs. Our measured $\sigma_{\rm int,H\alpha}$ and $\sigma_{\rm int,UV}$ across the full mass range at $1.4<z\leq 2.7$ agree with the predictions from FIRE at $z=2$.

For IllustrisTNG, \citet{2019MNRAS.485.4817D} reports that in the stellar mass range $10^9\text{--}10^{10.5}\ M_\odot$, their 100 cMpc simulation predicts a scatter of 0.2 dex for a 200-Myr-timescale SFR indicator (which we assume closely matches the UV luminosity) and 0.25 dex for an instantaneous SFR indicator (which we assume closely matches the H$\alpha$ luminosity). However, our measured scatter in both the H$\alpha$-based and UV-based SFMS exceeds the predictions made by the IllustrisTNG simulations at $z=2$. 

The disagreement of our observations with IllustrisTNG may come down to the simulation's treatment of star formation and feedback that are on scales too small to resolve. These processes are thought to contribute to the stochasticity of star formation \citep[e.g.,][]{2018MNRAS.480..800H}, and the inability to resolve these processes may affect predictions regarding the burstiness of SFHs. Considering this, our observation that $\sigma_{\rm int,H\alpha} > \sigma_{\rm int,UV}$ out to $z\sim 7$ is consistent with the interpretation that SFHs are bursty. The caveat to this, of course, is that the systematic uncertainties associated with converting H$\alpha$ flux to SFR are sub-dominant with respect to fluctuations in the galaxy SFHs. We discuss this caveat further in \ref{sec:confounding_factors}.

Whether the scatter in the SFMS varies as a function of stellar mass at high redshift does not have a clear consensus among theoretical works. \citet{2015MNRAS.451..839D} found that the scatter in H$\alpha$ luminosity varies as a function of stellar mass, with a scatter of 1.0 dex at the low-mass end ($10^7\ \rm M_{\odot}$) and 0.15 dex at the high-mass end ($10^{9.5}\ \rm M_{\odot}$). Several other theoretical works also find a mass dependence in the SFMS scatter \citep[e.g.,][]{2018MNRAS.478.1694M,2018MNRAS.480.4842C, 2019ApJ...879...11K,2022MNRAS.509..595L}, while some studies find no such trend \citep[e.g.,][]{2019MNRAS.485.4817D,2023A&A...677L...4P}. In our galaxy sample, we do not see any consistent mass dependence on $\sigma_{\rm int}$ in the SFMS between the low- and high-mass bins across redshift.

Since there is no consensus among theoretical predictions of the stochasticity of star formation or its mass dependence, observational constraints are critical to our understanding moving forward. The fact that our sample shows no consistent mass dependence of the scatter in the SFMS is in tension with theoretical works that do predict a mass dependence. This discrepancy suggests that star formation and feedback implementations in simulations and models that produce a mass-dependent scatter need to be revised. This, however, assumes that our galaxy sample spans a large enough dynamic range in mass, has negligible systematic uncertainties, and is large enough to probe mass-dependent scatter, which may not be the case. Future observational works with large samples of galaxies are needed to more robustly rule out theoretical implementations of star formation and feedback.

\subsubsection{Comparison to observational works}

In addition to theoretical predictions, previous observational works have constrained $\sigma_{\rm int}$ in the SFMS. Studies that use SFR indicators tracing star formation on 10-Myr timescales typically find intrinsic scatter values of $\sigma_{\rm int} \sim 0.2\text{--}0.5$ dex \citep[e.g.,][]{2013ApJ...777L...8K,2015ApJ...815...98S, 2023arXiv231210152C,2024ApJ...972..156N}. Many other studies find an intrinsic scatter of 0.2--0.3 dex when analyzing SFR indicators that trace star formation over longer timescales \citep[e.g.][]{2012ApJ...754L..29W,2015A&A...575A..74S,2016ApJ...817..118T} with some studies finding a scatter up to 0.5 dex \citep[e.g.,][]{2022ApJ...936..165L}. 

\begin{figure}
    \centering
    \includegraphics[width=8.5cm]{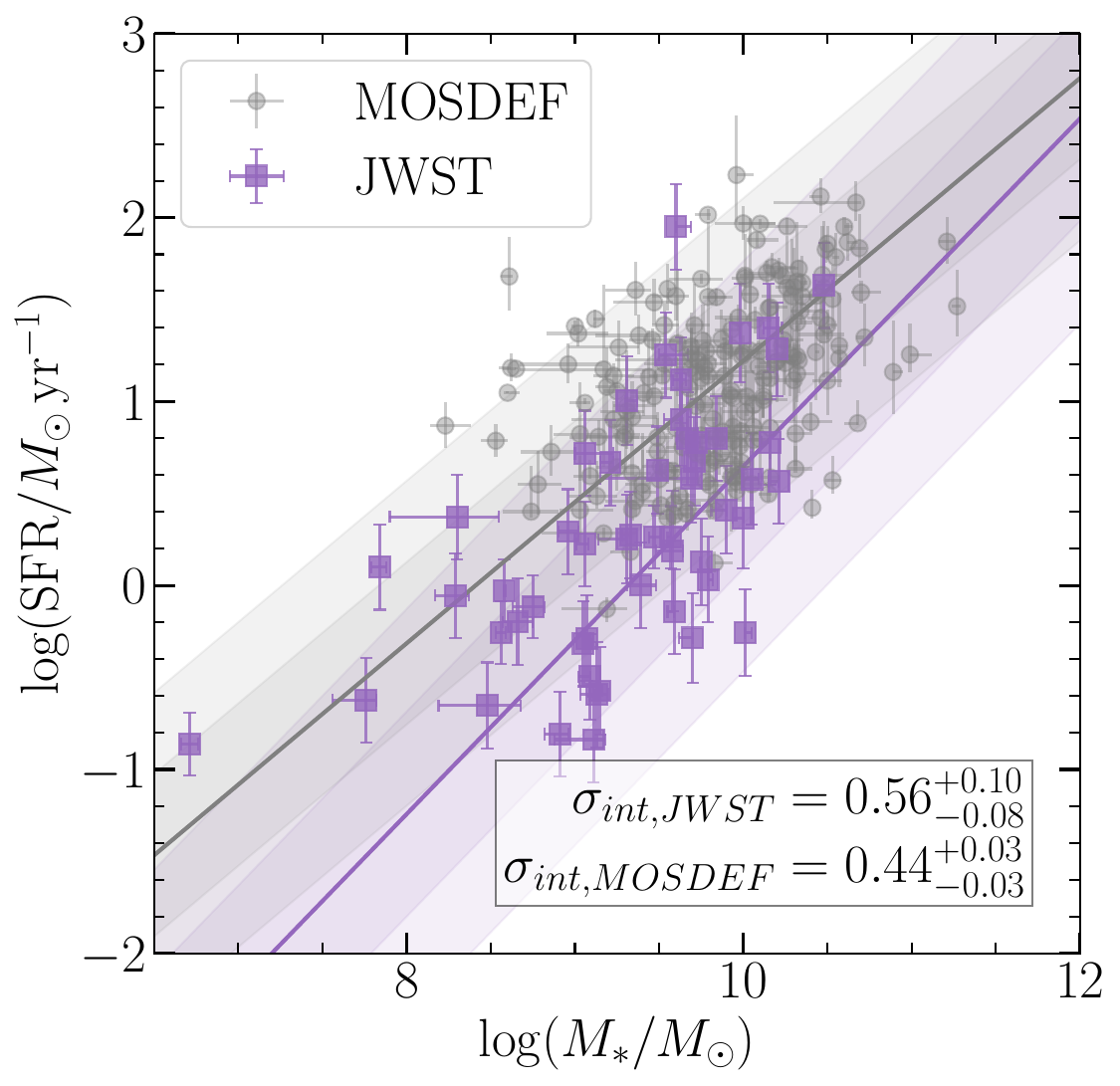}
    \caption{The H$\alpha$-based star-forming main sequence as determined from JWST observations in this study in the range $1.4 < z \leq 2.7$ ({\it purple}) and MOSDEF observations from \citet{2021ApJ...914...19S} in the range $2.0 < z < 2.6$ ({\it gray}). The MOSDEF data points have been shifted down by 0.32 dex to match the H$\alpha$ luminosity to SFR conversion that we use in this paper.}
    \label{fig:MOSDEF_JWST}
\end{figure}

In the case of our measurements of $\sigma_{\rm int}$ from both UV-based and H$\alpha$-based SFRs, we find results that are consistent with previous literature. In Figure~\ref{fig:MOSDEF_JWST}, we compare our results to those found by \citet{2021ApJ...914...19S} in the MOSDEF survey, which uses an expanded version of the sample from \citet{2015ApJ...815...98S}. We note that our measured $\sigma_{\rm int, H\alpha}$ exceeds that found from the MOSDEF survey in a similar redshift range. However, our sample is more sensitive to lower H$\alpha$ luminosities and lower stellar masses. Thus, our interpretation is that including galaxies in this new parameter space increases the scatter when compared with previous work \citep{2015ApJ...815...98S}.

We note that our selection of star-forming galaxies based on the specific SFR may result in slight differences when comparing to other works that use other selection criteria such as color cuts \citep[e.g.,][]{2009ApJ...691.1879W,2009A&A...501...15F,2011ApJ...735...86W,2012ApJ...754L..29W,2024ApJ...961...73N}. The selection criterion that is chosen to identify star-forming galaxies has been shown to introduce 0.2 and 0.5 dex uncertainties in the SFMS scatter and normalization, respectively \citep[e.g.][]{2022ApJ...936..165L, 2024ApJ...972..156N}. However, our sSFR cut only removes a single galaxy from our sample, and the differences in SFMS properties based on star-forming galaxy selection are expected to be most pronounced at masses above $10^{10.5}$ M$_\odot$, which exceeds the masses of the majority of our sample. Thus, our selection of star-forming galaxies likely does not result in significant systematic differences between this work and others in the literature.

 As is the case in theoretical work, observational studies yield no consistent answer regarding the mass-dependence of the scatter in the SFMS. Some studies report an increase in scatter with increasing stellar mass \citep[e.g.,][]{2013ApJ...778...23G,2019MNRAS.483.3213P,2019MNRAS.490.5285P,2021MNRAS.505..947S,2023arXiv231210152C}, some do not report any monotonic mass-dependence on the scatter \citep[e.g.,][]{2007ApJ...660L..43N,2015A&A...575A..74S, 2015ApJ...815...98S, 2022MNRAS.509.4392D}, and some report a decrease in scatter with increasing stellar mass \citep[e.g.,][]{2015MNRAS.449..820W}. 

 When comparing our low-redshift sample with the MOSDEF \citep{2015ApJS..218...15K,2015ApJ...815...98S} sample, however, we tentatively observe a mass dependence. The median stellar mass of our JWST sample in the lowest redshift bin is $10^{9.48}\ M_\odot$, while the median stellar mass of the MOSDEF sample spanning $2.0<z<2.6$ is $10^{9.88}\ M_\odot$. The fact that these samples span different ranges in stellar mass and exhibit different amounts of scatter in the SFMS provide tentative evidence for mass-dependence of $\sigma_{\rm int}$ when probing a large enough range in $M_*$. A larger, mass-complete spectroscopic sample is needed, however, to make a definitive statement about the presence or absence of mass dependence on the intrinsic SFMS scatter.

We also note that we do not measure any strong redshift evolution in $\sigma_{\rm int}$ for any of the SFR indicators, as illustrated in Figure \ref{fig:sig_int_z}. There has been increased interest in determining whether an evolution in bursty SFHs can explain the excess of UV-bright galaxies at $z>10$ \citep[e.g.,][]{2023MNRAS.526.2665S,2023ApJ...955L..35S,2023MNRAS.525.3254S, 2023MNRAS.521..497M}. Though our sample does not extend past $z\sim 7$, a larger, higher-redshift sample with smaller measurement uncertainties on all of the SFR indicators would allow for a better determination of the evolution of $\sigma_{\rm int}$ with cosmic time.

\subsection{The H$\alpha$-to-UV luminosity ratio}

\begin{figure*}
    \centering
    \includegraphics[width=7in]{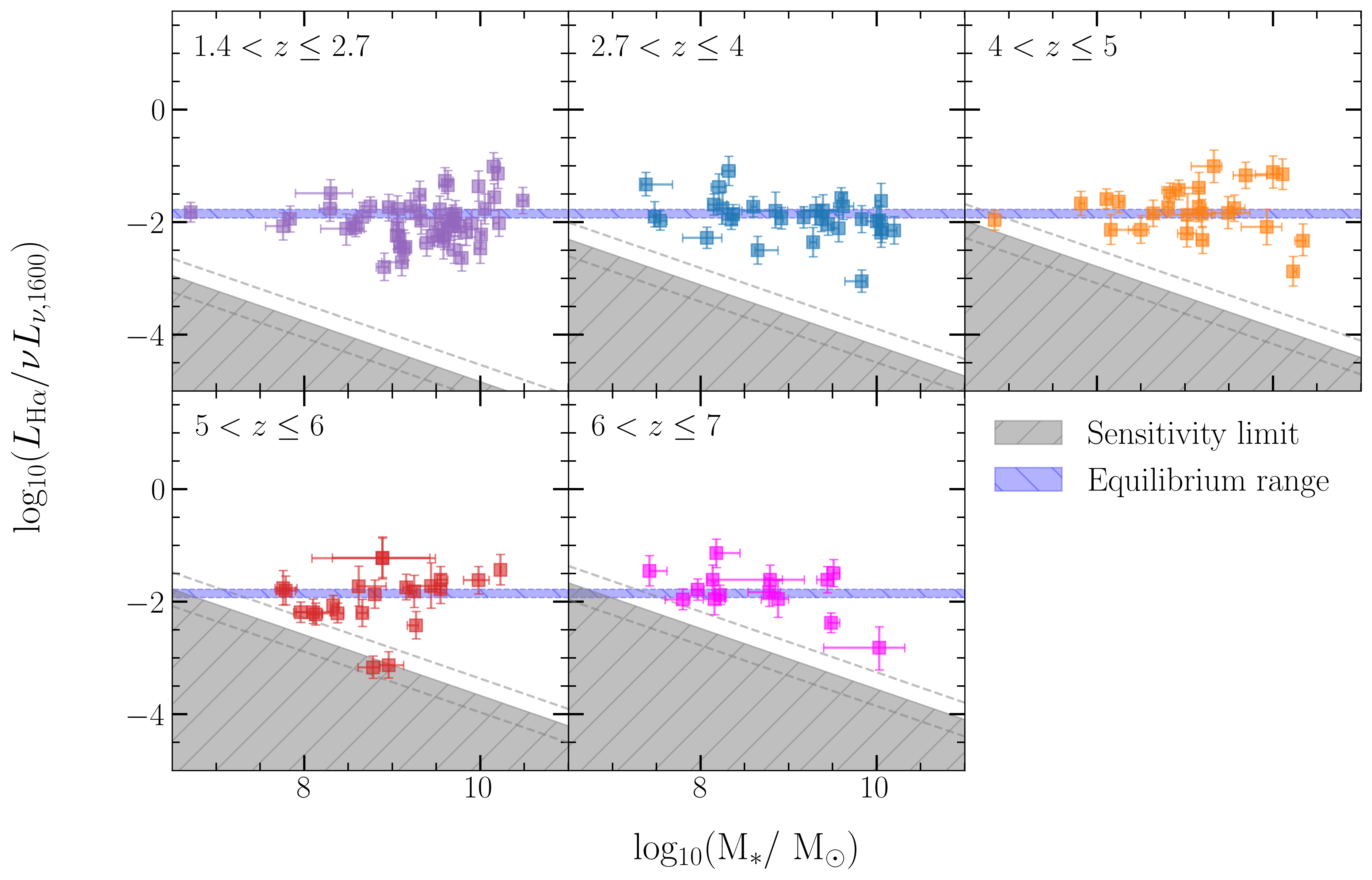}
    \caption{The H$\alpha$-to-UV luminosity ratio vs. stellar mass for our galaxy sample. The gray hatched region shows the parameter space that we are not sensitive to assuming a limiting survey flux and a UV-luminosity to stellar mass ratio (which we derive from the parameters reported in Table \ref{tab:fitparams}). We take the limiting flux in our sample to be $2.0\times 10^{-19}\ \rm erg\ s^{-1}\ cm^{-2}$ since this is the lowest H$\alpha$ flux we observe at a significance greater than 5$\sigma$. The dashed gray lines denote the uncertainty on the region of our sensitivity limits due to the intrinsic scatter in SFR(UV) vs. stellar mass relation. The horizontal blue hatched region shows the equilibrium range of H$\alpha$-to-UV luminosity calculated by \citet{2023ApJ...952..133M}.}
    \label{fig:ha_to_uv}
\end{figure*}

\begin{deluxetable}{c|cccc}
    \label{tab:ha_uv_ratio}
    \caption{The fractions of galaxies in vs. out of equilibrium SFHs in each redshift bin.}
    \tablehead{
    \colhead{Redshift Bin} & \colhead{$f_{eq}^{\tablenotemark{\scriptsize a}}$} & \colhead{$f_{above}^{\tablenotemark{\scriptsize b}}$} & \colhead{$f_{below}^{\tablenotemark{\scriptsize c}}$} & \colhead{$\sigma_{\rm H\alpha/UV}^{\tablenotemark{\scriptsize d}}$}
    }
    \startdata
    $1.4 < z \leq 2.7$ & 0.54 & 0.13 & 0.33 & $0.3^{+0.05}_{-0.05}$ \\
$2.7 < z \leq 4$ & 0.71 & 0.13 & 0.16 & $0.3^{+0.07}_{-0.06}$ \\
$4 < z \leq 5$ & 0.52 & 0.32 & 0.16 & $0.36^{+0.09}_{-0.08}$ \\
$5 < z \leq 6$ & 0.48 & 0.17 & 0.35 & $0.46^{+0.1}_{-0.09}$ \\
$6 < z \leq 7$ & 0.64 & 0.21 & 0.14 & $0.29^{+0.13}_{-0.1}$ \\
    \enddata
    \tablenotetext{a}{Fraction of galaxies whose H$\alpha$-to-UV ratio is consistent with the equilibrium range from \citet{2023ApJ...952..133M}.}
    \tablenotetext{b}{Fraction of galaxies whose H$\alpha$-to-UV ratio lies more than 1$\sigma$ above the equilibrium.}
    \tablenotetext{c}{Fraction of galaxies whose H$\alpha$-to-UV ratio lies more than 1$\sigma$ below the equilibrium.}
    \tablenotetext{d}{The measurement-subtracted scatter of the logarithmic H$\alpha$-to-UV ratio of the galaxies in each redshift bin.}
\end{deluxetable}

Another common metric used to characterize the burstiness of SFHs is the L(H$\alpha$)/$\nu \rm L_{\nu,1600}$ ratio. Since the H$\alpha$ and UV luminosities respond to changes in the SFR on different timescales ($\sim$5 Myr and $\sim$100 Myr, respectively), measuring their ratio gives information about the nature of SFHs for a galaxy population \citep[see][for more detailed discussions]{2021MNRAS.501.4812F, 2023MNRAS.526.1512R}. Using stellar population synthesis models and a constant SFH, \citet{2023ApJ...952..133M} found that the $\log({\rm L(H\alpha)/\nu \rm L_{\nu,1600}})$ ratio reaches an equilibrium after $\sim$100-200 Myr, reaching values the range of $-1.93$ and $-1.78$ for metallicities ($\log{(Z/Z_\odot)}$) between $-2$ and 0. Using this equilibrium range, \citet{2024MNRAS.52711372A} analyzed the H$\alpha$-to-UV luminosity ratio in their sample of $z \sim 4.7\text{--}6.5$ galaxies, finding that 60\% of their galaxies deviate from a constant SFH by greater than $1 \sigma$.

We plot the L(H$\alpha$)/$\nu \rm L_{\nu,1600}$ ratio of our galaxy sample in Figure \ref{fig:ha_to_uv}, along with the equilibrium range calculated by \citet{2023ApJ...952..133M} indicated by a blue, hatched, horizontal band. We additionally shade in the regions below which we do not expect to detect galaxies due to the sensitivity limits of the NIRSpec spectra ($\sim$$2\times 10^{-19}\ \rm erg\ s^{-1}\ cm^{-2}$). We convert this limiting flux to a lower limit on the detectable H$\alpha$ luminosity at the median redshift of the galaxy sample in each redshift bin. We then calculate the predicted value of $\nu L_{\nu,1600}$ as a function of stellar mass by translating our SFR(UV) vs. stellar mass linear fit parameters in Table \ref{tab:fitparams} to a relationship between $\nu L_{\nu,1600}$ and stellar mass. This results in a curve that represents our predicted sensitivity to $L_{\rm H\alpha}/\nu L_{\nu,1600}$ as a function of stellar mass.

In Table \ref{tab:ha_uv_ratio}, for each redshift bin, we show the fraction of galaxies whose H$\alpha$-to-UV ratios are consistent with the equilibrium value to within 1$\sigma$ ($f_{eq}$) and the fraction of galaxies with H$\alpha$-to-UV ratios above and below the equilibrium value ($f_{above}$ and $f_{below}$, respectively). We also calculate the standard deviation in the H$\alpha$-to-UV ratio, $\sigma_{\rm H\alpha/UV}$. Across redshift bins, we see that between 48\%--71\% of galaxies fall within the predicted equilibrium range of H$\alpha$-to-UV ratios. However, our measured $f_{eq}$ represents an upper limit. Galaxies that are transitioning between a period of decreasing star-formation (low H$\alpha$/UV ratio) and rising star-formation (high H$\alpha$/UV ratio) can be observed in the ``equilibrium" H$\alpha$/UV range due to time fluctuations in their SFHs rather than due to a constant SFH in the past 100 Myr. Thus, the observed number of galaxies in this ``equilibrium" range represents an upper limit on the number of galaxies which have actually been forming stars with a smooth SFH. We note that as a result, measuring the H$\alpha$/UV ratio in any individual galaxy is not sufficient to differentiate between burstiness or a smooth rising/falling SFH. However, we observe a $\sim$0.3 dex spread of log(H$\alpha$/UV) ratios at fixed stellar mass across redshift, suggesting that galaxies at similar evolutionary stages have a mixture of both rising and falling SFHs, rather than systematically being observed in a rising or falling SFH mode. We thus interpret this spread, in conjunction with $f_{eq}$, as being consistent with burstiness.

We also consider $f_{above}$ and $f_{below}$ in an attempt to constrain the duty cycle of star formation in the early universe. We expect that a low duty cycle of star formation will result in $f_{below} > f_{above}$ and vice versa. Our data reveal no consistent trend with $f_{above}$ or $f_{below}$ vs. redshift. In the lowest two redshift bins, $f_{below} > f_{above}$, suggesting a low duty cycle of star formation in this redshift range. However, $f_{above} > f_{below}$ in the $4<z\leq 5$ bin, which would suggest a high duty cycle. \citet{2023arXiv230605295E} suggested in their analysis of $z\sim 6$ galaxies that UV-faint (and thus low-mass) galaxies may be more likely to be observed in recent downturns in their star formation. This would produce low H$\alpha$ to UV ratios at low stellar masses in our sample in the high redshift bins. Due to the relatively low numbers of galaxies in these highest two redshift bins, however, it is difficult to ascertain how physically meaningful $f_{above}$ and $f_{below}$ are. Additionally, the sensitivity limits of the surveys begin to hinder our ability to measure $\log(\rm L_{H\alpha}/\nu L_{\nu, 1600})$ for low-mass galaxies at these redshifts. A future deep spectroscopic analysis of $z\sim 6$ galaxies would be able to provide a better view of the star-forming behavior of galaxies at this epoch. The models presented by \citet{2023ApJ...952..133M} are based on FSPS \citep{2009ApJ...699..486C,2010ApJ...712..833C}, and they do not account for binary stellar populations. \citet{2016MNRAS.456..485S} have shown that when normalized to 1500 angstroms, a stellar population of fixed age and metallicity will have a brighter ionizing continuum by a factor of $\sim$1.3 when assuming a model with vs. without binary stars. This would result in a 0.11-dex offset of the predicted “equilibrium” log(H$\alpha$/UV) ratio towards higher values. Consequently, our values of $f_{below}$ and $f_{above}$ would increase and decrease, respectively. We thus emphasize the importance of considering the effects of various SPS models when comparing the H$\alpha$/UV ratios to theoretical predictions, and the fact that this comparison is model-dependent.

Finally, we tabulate the measurement-subtracted scatter in the $\log(\rm L_{H\alpha}/\nu L_{\nu, 1600})$ distribution for each redshift bin. The scatter ranges between 0.30--0.46 dex, suggesting that time-dependent fluctuations in galaxy SFHs may result in differences in SFR that vary by a factor of $\sim$2--3 when determined from H$\alpha$ vs. UV luminosity.

\subsection{Important considerations}\label{sec:confounding_factors}

We have based our discussion thus far upon measurements of galaxy SFRs derived from H$\alpha$ and UV luminosities. There are, however, systematic uncertainties in these conversions that are difficult to quantify, and they may not be negligible.

The conversion between H$\alpha$ luminosity to SFR, for example, introduces some uncertainties. Deriving a SFR from hydrogen recombination line luminosities assumes that the escape fraction of ionizing radiation from galaxies ($f_{esc}$) is zero, which may not be the case in the early universe, especially for faint galaxies which are thought to be key contributors to cosmic Reionization \citep[e.g.,][]{2023MNRAS.524.2312E}. Direct measurements of $f_{esc}$ at $z\sim 3$ by \citet{2021MNRAS.505.2447P} suggest that galaxies typically have escape fractions of 5--10\%. Indirect studies of the SEDs of galaxies at higher redshifts suggest a range of escape fractions, with typical values of $\sim 13$\%  \citep{2024A&A...685A...3M},  reaching up to 50\% in the most extreme cases \citep{2022ApJ...941..153T,2023MNRAS.524.2312E}. However, the highest inferred escape fractions are characteristic of galaxies with the faintest UV luminosities ($M_{UV}\sim -17.5$) and bluest UV slopes ($\beta\sim -3$), properties not representative of the majority of our sample.  Thus, it is unlikely that uncertainties on $f_{esc}$ dominate the systematics in this study.

The conversion factor that we choose to translate betwen H$\alpha$ luminosity and SFR also introduces systematic uncertainties. We choose these conversion factors self-consistently with the metallicity assumptions used to derive stellar population properties from the SED fitting. However, calculating the ionizing photon production based on the H$\alpha$ luminosity likely involves more nuance than what is captured by choosing between two different metallicity-dependent conversion factors. The difference between the two conversion factors that we use is 0.22 dex, so the magnitude of the systematic uncertainty on SFR(H$\alpha$) is likely on this order. In addition, the fact that we use a single UV luminosity to SFR conversion factor rather than accounting for metallicity dependence introduces some uncertainty.

There are also uncertainties involved in choosing which dust attenuation law to apply to the stellar component of each galaxy in the sample. Several studies analyzing the infrared excess vs. UV slope relation in high redshift galaxies have determined that higher redshift, lower metallicity galaxies are better described by an SMC attenuation law, while local, metal-rich galaxies are better described by a \citet{2000ApJ...533..682C} dust law \citep[e.g.,][]{2018ApJ...853...56R}. However, though this trend may apply on average to galaxy populations, the specific dust law that best applies to an individual galaxy is difficult to determine, and assuming that a particular dust law applies to many galaxies may lead to increased scatter in the inferred SFRs \citep[e.g.,][]{2015ApJ...815...98S,2018ApJ...855...42S, 2024arXiv240805273S}. Within our sample, for the galaxies with $>$3$\sigma$ detections of H$\alpha$, H$\beta$, and H$\gamma$, we find that the median difference in E(B-V) when using all three lines vs. just H$\alpha$ and H$\beta$ is $0.00\pm 0.04$, suggesting that deviations from a \citet{1989ApJ...345..245C} nebular attenuation law do not dominate the systematic uncertainty in our sample.

We also note the possibility of missing nebular emission coming from optically-thick dust-enshrouded star-forming regions \citep{2012ARA&A..50..531K}. However, observations of Paschen lines by \citet{2023ApJ...948...83R} in galaxies at $z\sim 1\text{--}3$ suggest that this effect would boost the SFR(H$\alpha$) by 25\% at most. Additionally, we do not expect this effect to be particularly pronounced since most of our sample lies at low stellar mass ($\log(\rm M_*/M_\odot) < 10$), where dust attenuation is less significant \citep[e.g.,][]{2017ApJ...850..208W,2018MNRAS.476.3991M}. Within our sample, the median H$\alpha$ attenuation $A_{\rm H \alpha}$ is $0.23\pm 0.37$, while the median UV stellar attenuation $A_{1600}$ is $1.60\pm 1.46$, demonstrating the relatively low impact of dust attenuation on our sample.

Another important consideration is our use of a parameric SFH when performing SED fitting. Though the use of the delayed-$\tau$ SFH model is common in the literature \citep{2015ApJS..218...15K}, it has been shown that when compared to results derived from non-parametric SFHs, parametric SFHs may result in systematic uncertainties up to 0.4 dex \citep[e.g.][]{2020ApJ...904...33L,2024ApJ...963...74W}. However, when fitting the galaxy SEDs with a non-parametric SFH using \texttt{PROSPECTOR} \citep{2021ApJS..254...22J}, we found that the median difference between the Prospector and FAST stellar masses was 0.1 dex with a standard deviation of 0.3 dex, demonstrating that the non-parametric and delayed-$\tau$ SFH models yield consistent masses in this analysis.

Finally, variations in the stellar IMF may contribute significantly to derived galaxy properties \citep[e.g.,][]{2024ApJ...963...74W}. Throughout this analysis, we assume a \citet{2003PASP..115..763C} IMF, however, observations of galaxy properties in the early universe have led some authors to entertain the possibility that significant variations in the IMF exist in the early universe \citep[e.g.,][]{2022MNRAS.510.5603K,2022MNRAS.514L...6P}. It is not feasible to determine the specific form of the stellar IMF in each galaxy \citep[see][for a more recent review on the subject]{2024arXiv240407301H}, but this may be a source of uncertainty in our measurements if the IMF for early galaxies does vary significantly from the \citet{2003PASP..115..763C} functional form.

\section{Conclusions} \label{sec:conclusion}

In this study, we have analyzed the galaxy SFMS at $1.4<z<7$ using a combined CEERS and JADES NIRSpec sample. We measure the SFRs using different indicators: H$\alpha$, SED-fitting output, and UV luminosity. With these various indicators, we are able to investigate the burstiness of galaxy SFHs in the early universe. Our main conclusions are as follows:

\begin{enumerate}
    \item We find that the intrinsic scatter in the H$\alpha$-based SFMS is larger than the UV-based SFMS by $0.21 \pm 0.05$ dex on average. This difference is consistent with the idea that galaxies with bursty SFHs will deviate from the SFMS on short timescales, resulting in increased scatter for short timescale indicators such as H$\alpha$ compared to longer timescale indicators such as the FUV luminosity.
    \item We do not see evidence for mass-dependence in the intrinsic scatter about the SFMS within our sample. We do, however, see tentative evidence for larger scatter at lower masses when comparing our sample with the MOSDEF survey, suggesting that a large dynamic range in stellar mass is required to probe any mass dependence in $\sigma_{\rm int}$. 
    \item We also find evidence for bursty SFHs in some galaxies when analyzing the H$\alpha$-to-UV luminosity ratio. Depending on the redshift, 29--52\% of galaxies in our sample deviate from the ratio predicted for a stellar population with a constant SFH over a 100 Myr timespan. The degree of scatter about this equilibrium ratio also suggests the influence of burstiness on the SFHs of the galaxies in our sample. In order to fully see how this ratio behaves as a function of stellar mass, a deep, mass-complete spectroscopic sample is required.
\end{enumerate}

This study represents a preliminary analysis of the characteristics of the SFHs of star-forming galaxies in the early universe with multiple SFR indicators. Evidence for bursty star-formation histories has important implications for several areas, namely in the determination of stellar population properties through SED fitting and matching of theoretical simulations of galaxy formation with observations. We anticipate the promise of repeating this analysis with order-of-magnitude larger samples at similar redshifts covered by multiple SFR indicators.

%% Also note that the akcnowlodgment environment does not support long amounts of text. If you have a lot of people and institutions to acknowledge, do not use this command. Instead, create a new \section{Acknowledgments}.

\section{Acknowledgements}
We acknowledge the CEERS and JADES teams for their effort to design, execute, and make public their observational surveys. We would also like to thank Kate Whitaker, Zo\"e Haggard, James Bullock and his research group, as well as our anonymous reviewers for useful discussions that improved the quality of this work. We also acknowledge support from NASA grant JWST-GO-01914. This work is based on observations made with the NASA/ESA/CSA James Webb Space Telescope. The data were obtained from the Mikulski Archive for Space Telescopes at the Space Telescope Science Institute, which is operated by the Association of Universities for Research in Astronomy, Inc., under NASA contract NAS5-03127 for JWST. Data were also obtained from the DAWN JWST Archive maintained by the Cosmic Dawn Center. The specific observations analyzed can be accessed via \dataset[DOI: 10.17909/z7p0-8481]{https://archive.stsci.edu/doi/resolve/resolve.html?doi=10.17909/z7p0-8481}, \dataset[DOI: 10.17909/8tdj-8n28]{https://archive.stsci.edu/doi/resolve/resolve.html?doi=10.17909/8tdj-8n28}, \dataset[DOI: 10.17909/gdyc-7g80]{https://archive.stsci.edu/doi/resolve/resolve.html?doi=10.17909/gdyc-7g80}, and \dataset[DOI: 10.17909/fsc4-dt61]{https://archive.stsci.edu/doi/resolve/resolve.html?doi=10.17909/fsc4-dt61}.

% \begin{acknowledgments}

% \end{acknowledgments}


\vspace{5mm}
\facilities{\it JWST, HST}


\software{ Astropy \citep{2013A&A...558A..33A,2018AJ....156..123A,2022ApJ...935..167A}, \texttt{emcee} \citep{2013PASP..125..306F}, \texttt{jwst} \citep{2023zndo...7577320B}, \texttt{msaexp} \citep{2022zndo...7299500B}, \texttt{scipy} \citep{2020NatMe..17..261V}}




\bibliography{main631}{}
\bibliographystyle{aasjournal}



\end{document}